\documentclass[10pt, a4paper]{article}

\usepackage{lrec-coling2024} 
\usepackage{xurl}
\usepackage{graphbox}
\usepackage{subcaption}
\usepackage{enumitem}
\usepackage{xcolor}

\title{Social Convos: Capturing Agendas and Emotions on Social Media} %

\name{Ankita Bhaumik, Ning Sa, Gregorios Katsios, Tomek Strzalkowski} 

\address{Rensselaer Polytechnic Institute\\
         Troy, New York \\
         \{bhauma, san2, katsig, tomek\}@rpi.edu\\}

\abstract{
Social media platforms are popular tools for disseminating targeted information during major public events like elections or pandemics. Systematic analysis of the message traffic can provide valuable insights into prevailing opinions and social dynamics among different segments of the population. We are specifically interested in influence spread, and in particular whether more deliberate \textit{influence operations} can be detected. 
However, filtering out the essential messages with telltale influence indicators from the extensive and often chaotic social media traffic is a major challenge.
In this paper we present a novel approach to extract influence indicators from messages circulating among groups of users discussing particular topics. We build upon the concept of a \textit{convo} to identify influential authors who are actively promoting some particular agenda around that topic within the group. We focus on two influence indicators: the (control of) agenda and the use of emotional language.
 \\ \newline \Keywords{agenda, emotions, social networks, campaigns} }

\begin{document}

\maketitleabstract

\section{Introduction}
\label{sec:intro}
Popular social media platforms, such as Twitter (now X), Facebook, Instagram, and others, have become global, wide reaching communication channels that remain largely unregulated for access and content. They are an ideal vehicle for placing precisely targeted commercial advertising and other types of outreach activities. Social media have also become means of choice for spreading propaganda, disinformation, and for running influence operations. In this paper we focus on this last phenomenon; more specifically: how the presence of deliberate influence operations can be detected among the overall message traffic. 

While at the first glance the social media message traffic may appear random and chaotic, it is in fact an environment where all kinds of social phenomena obtain, not unlike in any other forms of human interaction, except at a much larger scale. Social media is thus as an extensive repository of information to study group dynamics at scale, including sociolinguistic behaviors such as agenda control and influence, as well as the methods and techniques deployed to achieve them \citep{del2017mapping, bovet2019influence, caldarelli2020role}. 


At the same time, a crucial challenge for an effective analysis of the message traffic is separating specific interactions from among many that may be occurring at the same time, all set against the backdrop of largely unrelated "noise". If we want to analyze an potential influence operation, we must zero in on a subset of messages shared between the possible influencers and their targets, as well as many onlookers. It is within this potentially large group where the key social behaviors can be observed with sufficient clarity. Previous work has been done on the representation of such behaviors on social media as a landscape of \textit{convos} \citep{Katsios-2019}, and this paper builds upon it.

\citealp{Katsios-2019} define a \textit{convo} as an online social phenomenon where people are engaged around a topic or an activity. A \textit{convo} can be formed among users who read, edit, comment or forward an information artifact, which could be a document, a project, a topic, or an idea. Some examples can be a repository on GitHub, a subreddit on Reddit or a hashtag group on Twitter. We hypothesize that \textit{convos} can be used to detect focused groups of messages around a particular topic, which can be further used for detecting influence operations.


One well-documented situation where social media can be used to run influence operations is during national elections and public emergencies, such as the recent COVID-19 pandemic
\citep{karlsen2016styles, jungherr2016twitter, badawy2019characterizing}. These operations are driven by groups of users and characterized by the specific agendas they aim to promote. We deploy the concept of a \textit{convo} to extract the subsets of social media traffic devoted to popular topics arising within the specified time periods, e.g., during French national elections in 2022.
From the convo traffic, we extract the messages related to its main topic, as well as 
a network of influential authors who are actively promoting a particular viewpoint, or agenda within the convo. We characterize the \textit{convo} representation using well-established influence indicators, specifically, agenda control \citep{broadwell2013modeling} and use of emotional language\citep{bhaumik2023adapting}. This allows us to identify the most influential members of the convo, as well as to uncover a network of relationships between them. Our hypothesis is that a dense network of mutual connections among the key influencers in the convo indicates a deliberate influence operation. This may be contrasted with a more organic message traffic where individual influencers are acting largely independently of one another.

Existing works on extracting influence indicators from social media has been performed on a message level \citep{mather2022stance, bhaumik2023adapting}. After tagging individual messages with agenda and emotion labels, the aggregate agendas and emotions in the dataset are typically obtained by calculating a relative distribution or density of each label. However, this bottom-up approach has several drawbacks. Firstly, by working on individual messages, this approach underestimates the semantic and social links between the messages, thus treating them as independent events. 
Secondly, by ignoring duplicate messages and reposts (retweets on Twitter), this approach weighs each unique message equally and ignores the differences in their popularity. Thirdly, although there are predefined emotion categories, the set of agendas germane in each type of event are typically unknown in advance. As a result, some amount of human annotation may be needed in order to obtain the relevant agenda labels.

To mitigate the above drawbacks, we propose a novel top-down approach to detect agendas (and who pushed them) and emotions (and who uses them) simultaneously from focused groups of messages in a \textit{convo}. We use instruction tuned pre-trained large language models (LLMs) that have the ability to process collections of messages as opposed to fine-tuned models that work at a message level.

We perform our experiments on the 2022 French Election Twitter dataset to study popular influence operations, such as \#Frexit \citep{frexit}, that promoted France's exit from the European Union. We outline the complete methodology of extracting \textit{convos} from a social media corpus, identification of influential authors in a convo and characterizing them using agendas and emotions in Section \ref{sec:methodology}. Though our experiments have been carried out on a Twitter dataset, our approach has been generalized for multiple social media platforms like Reddit or public forums which do not use hashtags.

\section{Methodology}
\label{sec:methodology}

\begin{figure*}[!h]
\includegraphics[width=0.98\textwidth]{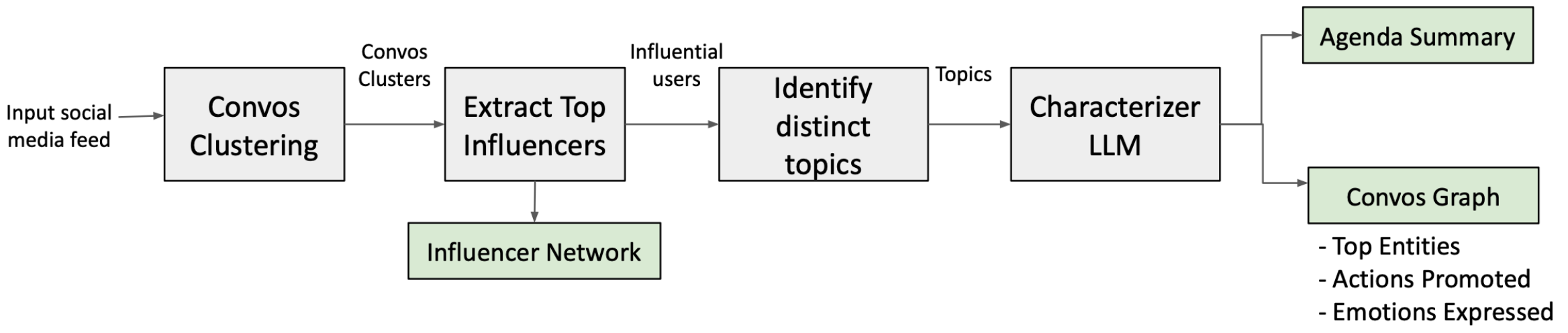}
\caption{Overview of our approach to extract and characterize \textit{convos}.}
\label{fig:workflow}
\centering
\end{figure*}

Our approach to identify \textit{convo} networks comprises of three major steps (Fig. \ref{fig:workflow}):
\begin{enumerate}[itemsep=0.05ex, topsep=1pt]
    \item Identification of a convo using hashtag or topic groups around the keywords of interest.
    \item Identification of authors who act as top influential users in a convo.
    \item Detection of entities, agendas and emotions from the convo.
\end{enumerate}

In the following subsections, we outline every step in detail. 

\subsection{Identification of Convos}
As described by \citeauthor{Katsios-2019}, the definition of a convo depends on the type of social media under examination. For example, on Twitter, convos typically form around hashtag groups and on Reddit they arise in subreddits discussing similar topics. Therefore, given a set of terms of interest, the first step in our approach is to identify these hashtag or topic groups in the input corpus.

Employing a keyword search mechanism to identify the hashtag or topic groups on social media is a less effective approach due to the inherent complexities of the medium. Social media messages are often short and may not mention certain keywords directly. For example, when examining tweets from the 2022 French Elections presidential campaign, its not guaranteed that the terms \texttt{France},  
 \texttt{president}, or \texttt{election} will be explicitly mentioned. To address this challenge, we first identify hashtag groups on Twitter, or topic groups in other social media platforms, and we use these groups to gather messages related to keywords of interest, allowing for a more comprehensive and accurate analysis.

To identify hashtag convos, we create a co-occurrence based distance matrix of the top 6000 tokens in the hashtag vocabulary of the corpus. Using this as our feature representation, we perform dimensionality reduction via UMAP \citep{2018arXivUMAP} and then use HDBSCAN \citep{campello2013density} to cluster them into hashtag groups. Some examples of hashtag groups are listed in Table \ref{tab:hashtag_groups}.

For datasets from other social media like Reddit or public forums, we use the LDA topic modeling algorithm to identify topic distributions in the input corpus \citep{blei2003latent}. Finally, we identify convos around a particular topic using the hashtag/topic groups containing any of the terms of interest.

\begin{table}[!htb]
    \centering
    \begin{tabular}{p{0.95\linewidth}} 
    \hline 
        \textbf{Hashtag Groups}\\ \hline 
        \#frexit, \#ue, \#asselineau, \#reprenonslecontrôle, \#schengen \\ 
        \hline 
        \#stopputin, \#stoprussia, \#standwithukraine, \#closetheskyoverukraine, \\
        \#noflyzoneoverukraine\\ 
        \hline
        \#covid19, \#covid, \#covid\_19, \#omicron, \\\#vaccine \\ 
        \hline 
        \#passvaccinal, \#passevaccinal, \#passsanitaire, \#passdelahonte, \#passevaccinaldelahonte\\ 
        \hline 
        \#mckinseygate, \#macronmckinsey, \#macronassassin, \#macronmckinseygate, \#macroncnon\\ 
        \hline
    \end{tabular}
    \caption{Examples of hashtag groups extracted from the French Elections 2022 dataset, showing top 5 hashtags in each group.}
    \label{tab:hashtag_groups}
\end{table}

\subsection{Identification of the Top Influencers}

As previously discussed, not all messages within a hashtag convo carry equal significance. The weight of a message is determined by the number of retweets, reflecting broader agreement among users. Conversely, a collection of messages from a single user can provide a more comprehensive expression of their ideas and positions compared to a single message. Therefore, instead of analyzing all tweets in a convo, we prioritize the examination of tweets from influential users, who are often retweeted.

To characterize these influential tweets we categorize them by topic, allowing us to extract the agendas and emotions associated with each topic group. We use a two-level hierarchical clustering method to identify distinct topics within bigger groups of messages \citep{grootendorst2022bertopic}. The agendas and emotions expressed by these influential users serve as a representative sample of the entire convo.

Furthermore, we explore the underlying efforts driving these agendas by scrutinizing the tweeting behavior of influential users and exploring the connections within the cross-influencer network. We hypothesize that in a targeted influence operation these authors would be actively retweeting each other to promote similar ideas.


\subsection{Agenda and Emotion Detection}

In order to effectively perform the tasks of agenda and emotion detection across a collection of messages, we utilize large language models (LLMs) as they have a large context window. Existing fine-tuned models are not suitable for these tasks, primarily due to their limitations in processing single messages within specific domains. Recent advancements in pre-trained LLMs have demonstrated remarkable capabilities in zero-shot summarization across diverse domains \citep{zhang2023benchmarking, yang2023exploring}. Leveraging these breakthroughs, we have developed an instruction-tuned model for agenda and emotion detection.

In our approach, we provide the LLM with a global context that acts as an instructional guide for its intended task. We adopt a prompt-based template to extract both the agendas and emotions from a collection of messages. Additionally, we employ an output template designed to steer the LLM in generating outputs in a format conducive to visualizing all components within the convo network. This output template replicates a JSON file structure, enhancing its utility for seamless downstream processing. We design prompts to direct the model to answer the questions and accomplish the following tasks: 
\begin{itemize}[itemsep=0.05ex, topsep=1pt]
    \item Prompt 1: \textit{What are the top distinct entities (maximum 5) mentioned in several messages?}
    \item Prompt 2: \textit{What are the authors promoting about each entity? Give me 1 phrase for each.}
    \item Prompt 3: \textit{What are the emotions expressed towards each entity?}
    \item Prompt 4: \textit{Combine the entities, promoted actions and emotions in the output template.}
\end{itemize}

The prompts are used to fill out an output JSON template:
\begin{verbatim}
output = [
{
"entity": {entity},
"promoted_actions": {action},
"emotions": {emotion}
},
...
]
\end{verbatim}

This yields a file that contains information regarding the entities, agendas, and emotions within the conversation.

\section{Experiments}

\subsection{Dataset}
We run our experiments on a publicly available dataset from the 2022 French presidential elections \cite{frenchElection}. It contains 45 million tweets and retweets from Nov. 12, 2021 to Apr. 03, 2022, around the main actors in the election. French is the dominant language of the dataset, and there are also certain amount of tweets written in English. This dataset contains all retweet information about the tweets which makes it an ideal choice to identify influential message groups in this corpus. We choose two keywords of interest that were popular topics during the French election campaign as studied from news articles and online forums : \#frexit and \#covid\_19 \citep{frexit}.

\begin{figure*}[!t]
\begin{subfigure}[t]{.45\textwidth}
  \centering
  \includegraphics[width=0.7\linewidth]{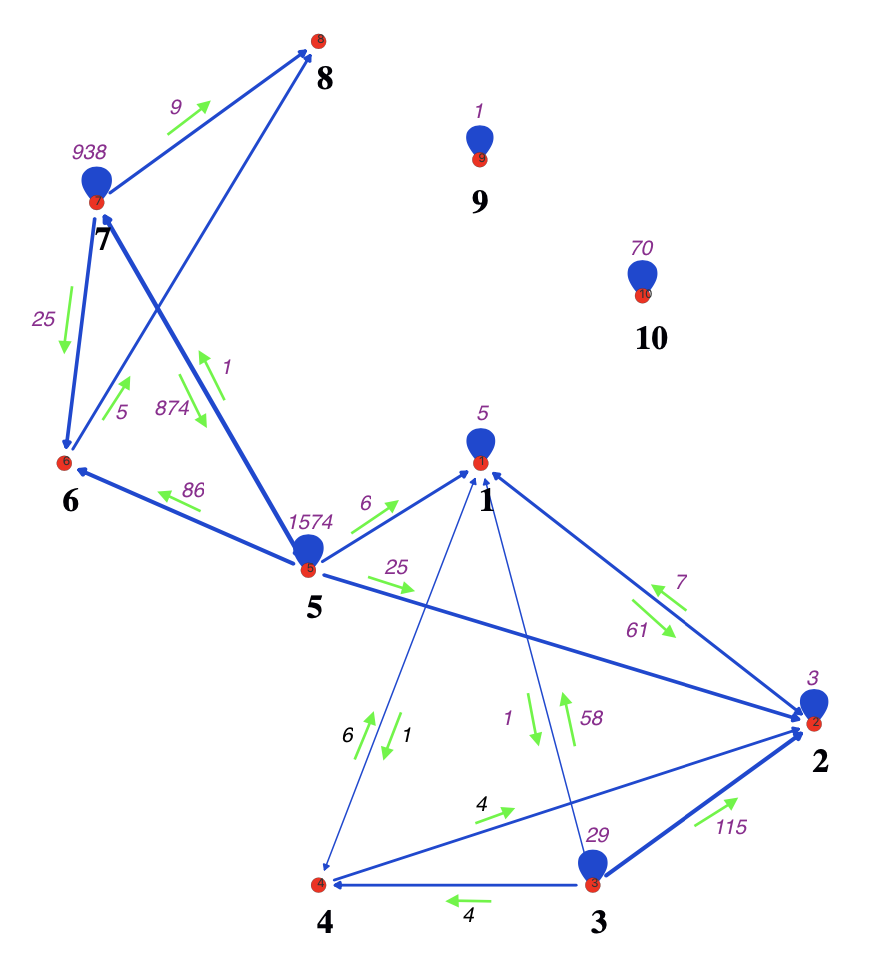} 
  \caption{\#Frexit convo}
  \label{fig:influencer_frexit}
\end{subfigure}
\begin{subfigure}[t]{.45\textwidth}
  \centering
  \includegraphics[width=0.8\linewidth]{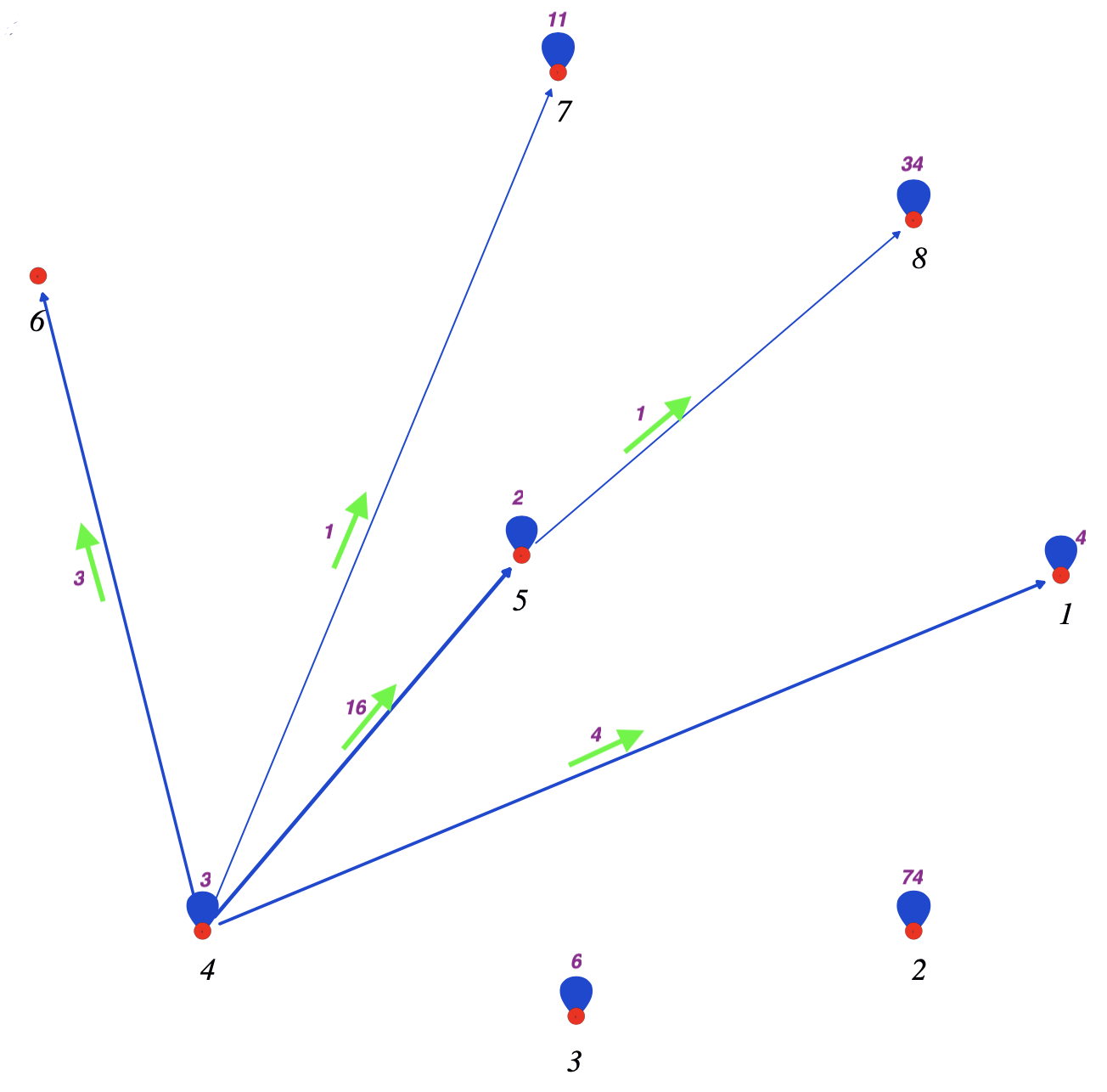}
  \caption{\#covid\_19 convo}
  \label{fig:influencer_covid}
\end{subfigure}

\caption{Influencer Networks among top influencers in each convos. Each node (red point) represents an influencer, with her index marked below the node. The arrows indicate the direction of the connections and the numbers above show the weight of the connections. A balloon on top of a node means a self-RT link.}
\label{fig:influencer_network}
    \end{figure*}

\subsection{Implementation}
To make a traceable retweet network, we pre-process the original dataset to link all retweets to the original tweet. We also remove emojis, hyperlinks and tweets having less than 3 textual tokens. To pick the most influential messages in a \textit{convo} we examine at the messages by the top 10 most retweeted authors in that \textit{convo}. Alternatively, this value could be determined based on the proportion of messages contributed by each author in a \textit{convo}. We use multilingual BERTopic to separate the \textit{convo} messages into smaller groups \citep{grootendorst2022bertopic}.

To implement the agenda and emotion detection system we use a LLM with a sufficiently large context window. We perform our experiments using Llama-2-13b-chat model with a context length of 4096 tokens \citep{touvron2023llama}. We provide the global context in the system prompt and use the prompts to further probe the LLM. 
We use nucleus sampling with $p$ as 0.9 for generating responses and limit the number of new tokens to 500. We also tested our approach using ChatGPT to achieve comparable performance \citep{ouyang2022training}. The experiments are performed on 2 NVIDIA A100 GPUs.

\subsection{Results}
To establish a mapping between each tweet and its corresponding retweets, we preprocess the original corpus, resulting in a dataset comprising of 16.7 million tweets. On further removal of emojis, hyperlinks, tweets without hashtags, and tweets having less than 3 textual tokens, we arrive at 2.8m tweets and their associated retweet information.

Our first step to identify all hashtag groups in the corpus results in more than 40 hashtag groups (Table \ref{tab:hashtag_groups}). We select the two that contain our terms of interest, \textit{frexit} and \textit{covid\_19}, to extract the hashtag \textit{convos} around them. Table \ref{tab:convo_stats} lists the total number of authors, tweets and retweets in each of these convos.
We then select the messages by the top 10 retweeted authors. In the \#frexit convo, the top influencers' messages account for 5\% of total messages in the convo and comprise around 44\% of the total retweets (20,135 out of 44,859). Whereas for the larger \#covid\_19 convo, that has a larger number of tweets and authors, the top influencers contribute 1\% of all tweets and their retweet number comprise 28\% of the total retweets. 

To understand the connections among the top 10 influencers, we construct a network based on their retweet behavior in the whole corpus (Fig. \ref{fig:influencer_network}). The nodes in Fig. \ref{fig:influencer_network} represent the authors and the weight of the edges represent the number of retweets between them, balloons on top of the node indicate instances of self-retweeting by that author.
We construct a graph for each convo and notice that there is no overlap in top influencers between the two, signifying that these authors are predominantly dedicated to promoting specific agendas. While the top 10 influencers in the \#Frexit convo (Fig.\ref{fig:influencer_frexit}) are connected by a dense network of retweet or self-retweet, the network for \#covid\_19 (Fig. \ref{fig:influencer_covid}) contains only 8 authors.

\begin{table}[!h]
\small
    \centering
    \begin{tabular}{l|cc|cc}
        \hline
        & \multicolumn{2}{|c|}{\#Frexit} & \multicolumn{2}{c}{\#Covid\_19} \\
        \hline
        & Inf.   & Convo & Inf.   & Convo\\

        \hline
        No. authors              & 10 & 3,572 & 10 & 10,239 \\
        \hline
        No. tweets               & 902 & 16,406 & 403 & 39,719 \\
        \hline
        No. RT count             & 20,135  & 44,859 & 38,144 & 133,457 \\

        \hline
    \end{tabular}
    \caption{Descriptive results of influencers (Inf.) in the \#Frexit and the \#Covid\_19 convo }
    \label{tab:convo_stats}
\end{table}

\begin{figure*}[!t]
\begin{subfigure}{\textwidth}
    \begin{subfigure}[t]{.5\textwidth}
      \centering
      \includegraphics[width=\linewidth]{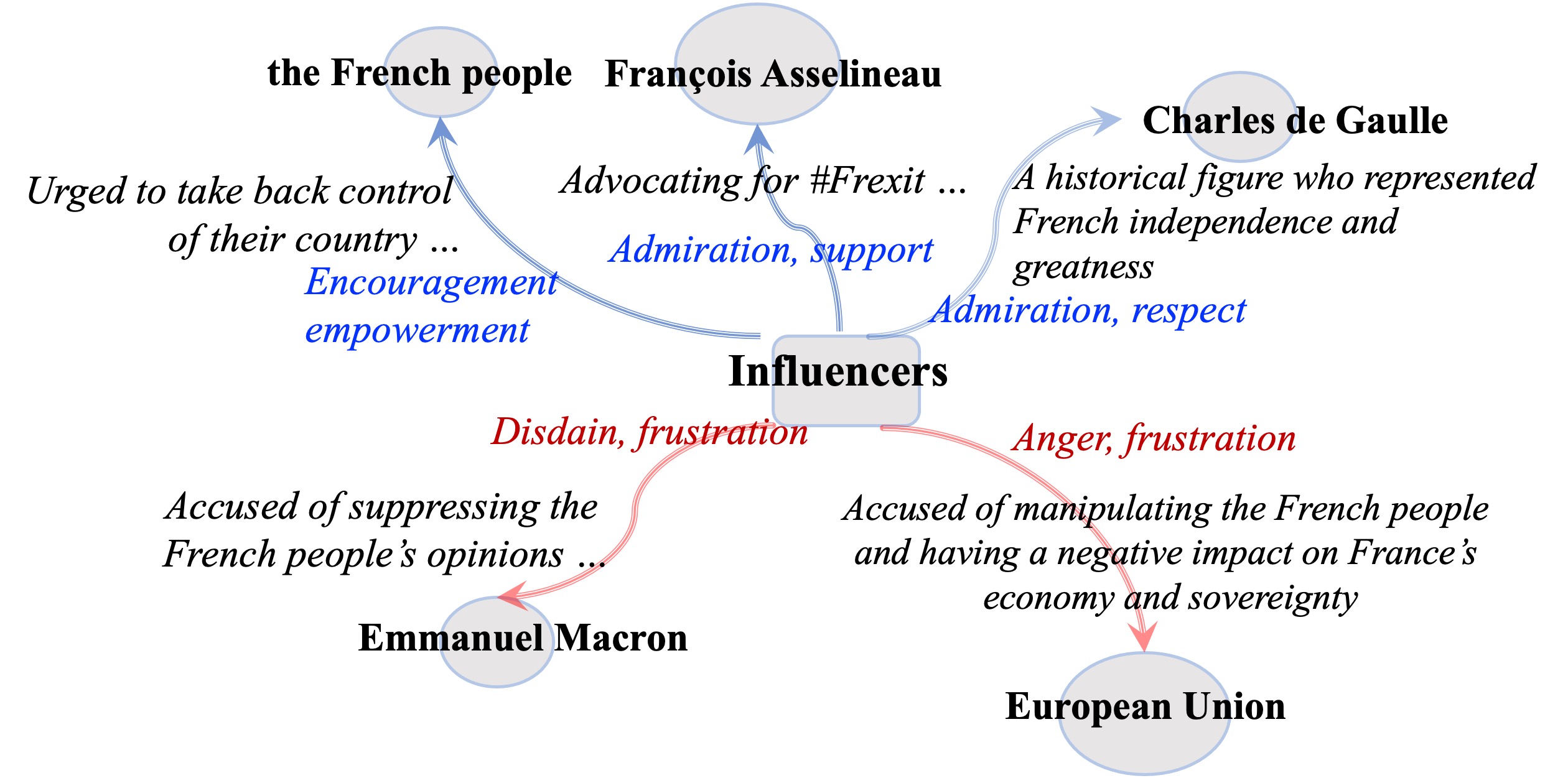}  
    \end{subfigure}
    \begin{subfigure}[t]{.49\textwidth}
      \centering
      \includegraphics[width=\linewidth]{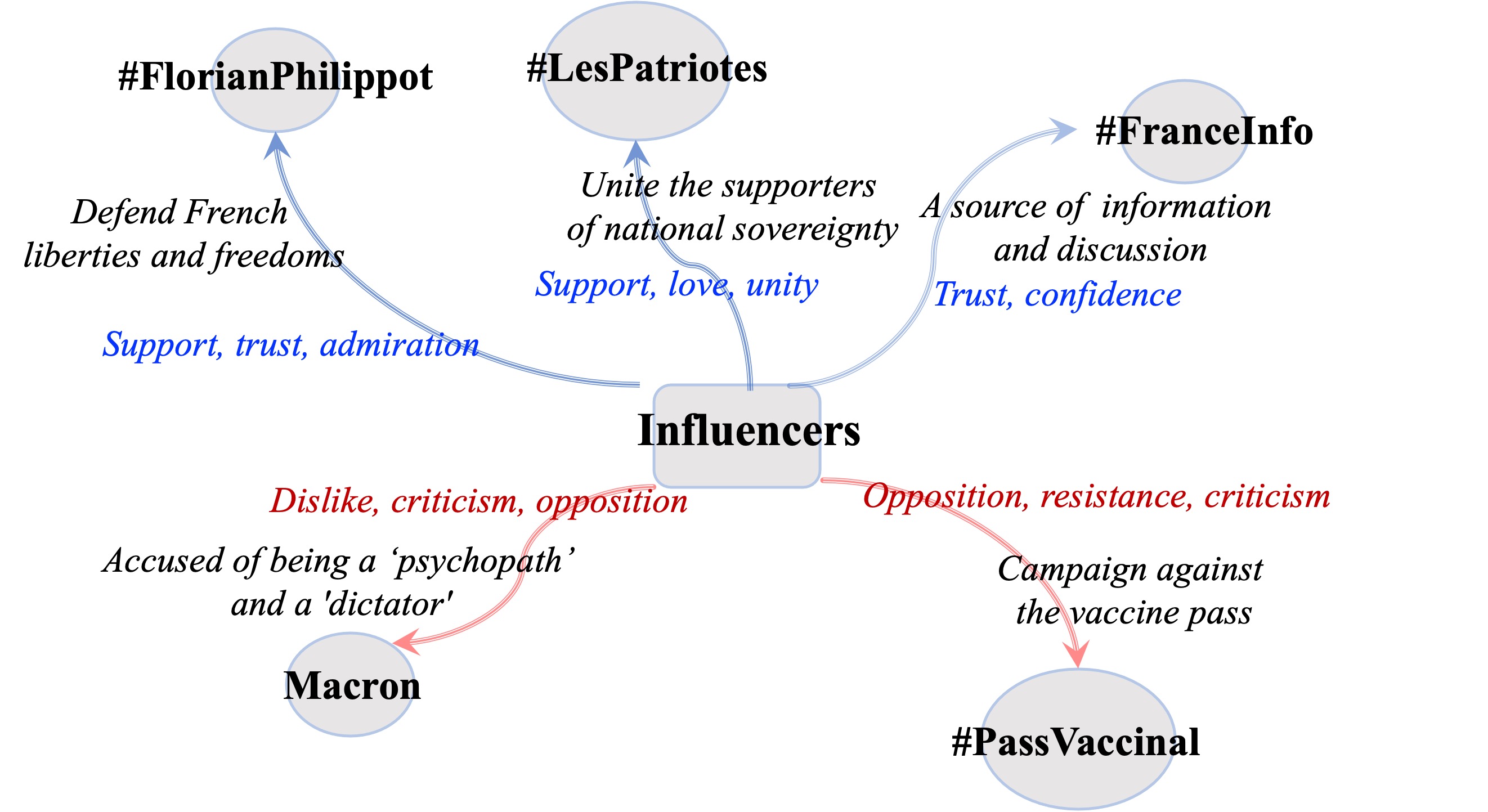}  
    \end{subfigure}
    \caption{Convos in the \#frexit clusters}
    \label{fig:frexit_clusters}
\end{subfigure}

\begin{subfigure}{\textwidth}
    \begin{subfigure}[t]{.49\textwidth}
      \centering
      \includegraphics[width=\linewidth]{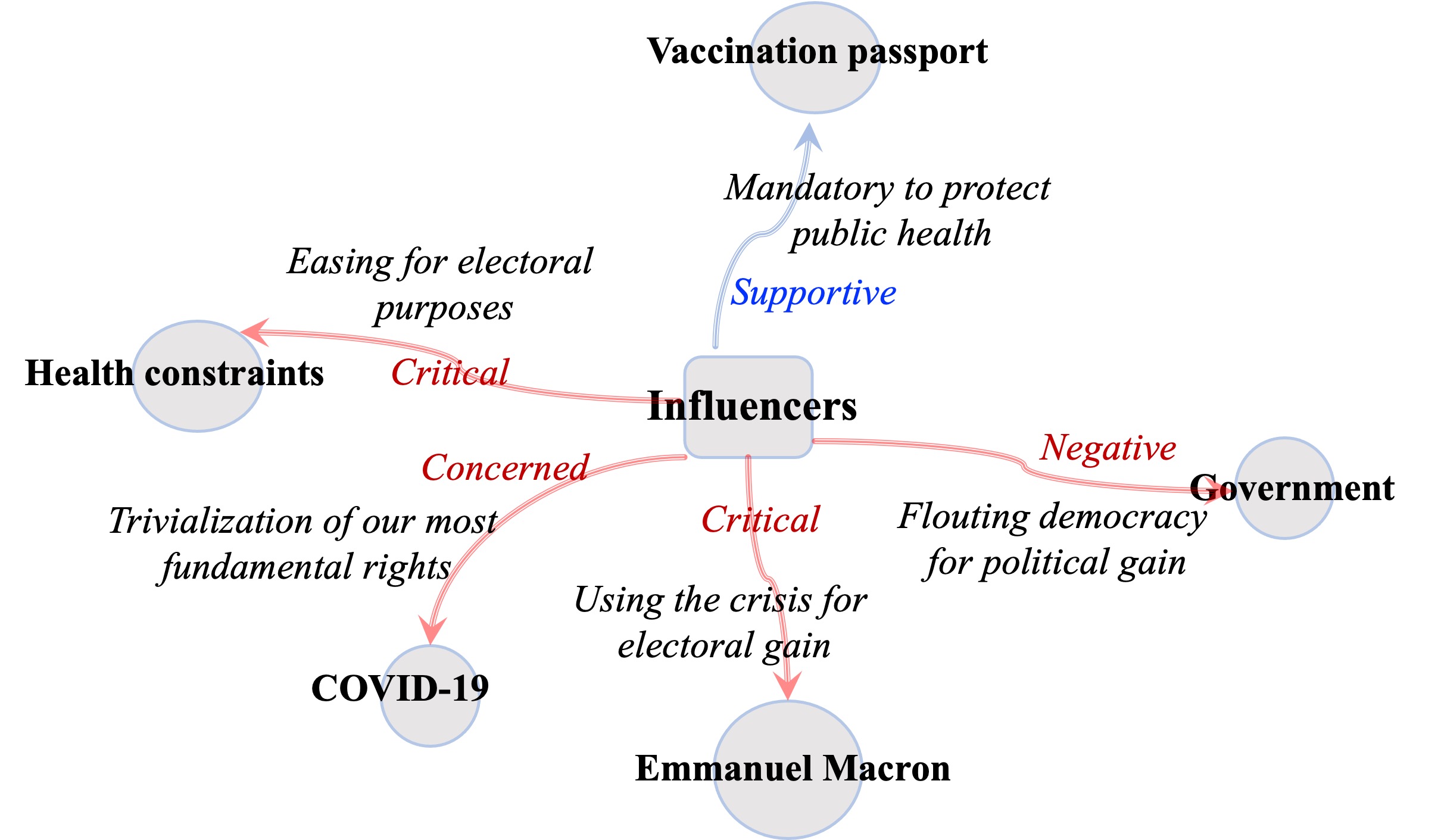}  
    \end{subfigure}
    \begin{subfigure}[t]{.5\textwidth}
      \centering
      \includegraphics[width=\linewidth]{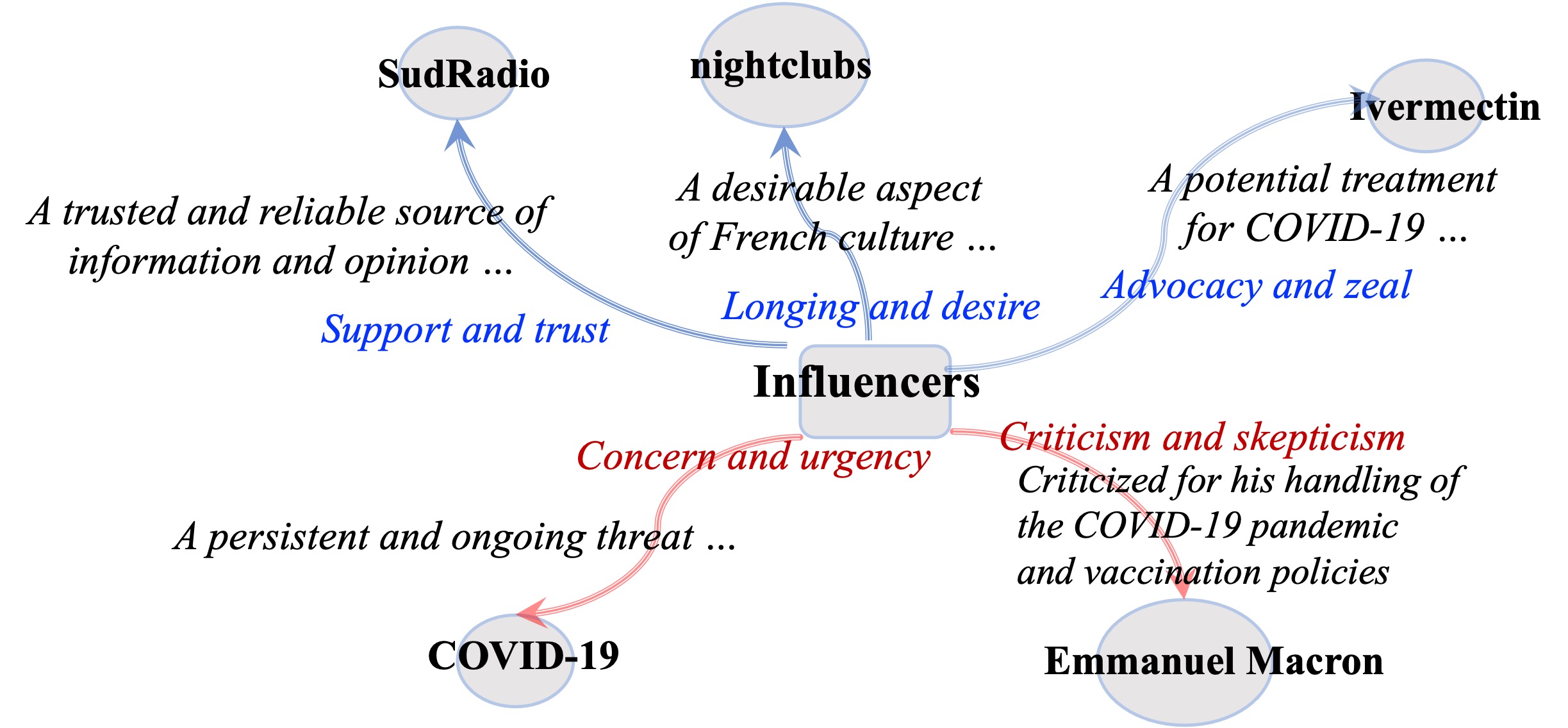}  
    \end{subfigure}
\caption{Convos in the \#covid\_19 clusters}
\label{fig:covid_clusters}
\end{subfigure}

\caption{Different clusters in the \textit{convos} about \#Frexit and \#covid\_19. They are characterized by the indicators of top entities, promoted actions, and emotions. The \textit{influencers} are the top 10 retweeted authors in that convo. The expanded network of the influencers are illustrated in Fig. \ref{fig:influencer_frexit} and \ref{fig:influencer_covid}.\\ 
Entities are grey blobs and promoted actions are the grey edge labels. Emotion labels over the edges are colored as \textcolor{red}{negative}/\textcolor{blue}{positive} ones. The model also generates an agenda summary, given in Table \ref{tab:agenda-summary}.}
\label{fig:convos_clusters}
\end{figure*}

\begin{table*}[!hbt]
    \centering
    \begin{tabular}{p{0.2\linewidth} | p{0.74\linewidth}} \hline 
         Message Cluster& Agenda Summary\\ \hline 
         Fig. \ref{fig:frexit_clusters} - left& The agenda behind this set of tweets is to promote the candidacy of François Asselineau for the French presidency in 2022 and to advocate for France's withdrawal from the European Union (\#Frexit)\\ \hline 
         Fig. \ref{fig:frexit_clusters} - right& The agenda behind this set of tweets is to promote the political movement \#LesPatriotes and its leader, Florian Philippot, and to campaign against the current French President Emmanuel Macron and his policies, particularly the controversial vaccine pass\\ \hline 
         Fig. \ref{fig:covid_clusters} - left& The agenda behind this set of tweets is primarily focused on the French government’s handling of the COVID-19 pandemic, specifically the use of vaccination passports, the easing  of health constraints, and the perceived political motivations behind these actions\\ \hline 
         Fig. \ref{fig:covid_clusters} - right& The agenda behind this set of tweets is to criticize and scrutinize the actions and words of French President Emmanuel Macron regarding the COVID-19 pandemic and vaccination\\ \hline
    \end{tabular}
    \caption{Summarized agendas of the messages clusters shown in Fig.\ref{fig:convos_clusters}}
    \label{tab:agenda-summary}
\end{table*}

We cluster the influencers' messages in each convo into smaller topic groups to be able to analyze the content more effectively. We run our emotion and agenda detection model over the message clusters to characterize the influence indicators in each influential convo. Fig \ref{fig:convos_clusters} visualizes the json outputs we obtain to depict the top entities, promoted actions and emotions of two selected message clusters from each convo. The generated agenda summaries, listed in Table \ref{tab:agenda-summary}, provide us with a concise and informative snapshot of the convos around the topic of interest. 

\paragraph{\#frexit convo (Fig \ref{fig:frexit_clusters}):}
In the first message cluster of this convo, the influencers promote the candidacy of François Asselineau and advocate for France leaving the EU. The first snapshot also illustrates a similar idea: the influencers support \textit{François Asselineau} who advocates for \#Frexit, encourage \textit{the French people} to take back control of their country, and admire \textit{Charles de Gaulle} who represented French independence and greatness. They also accuse \textit{Emmanuel Macron} for suppressing the French people's opinion and are angry about \textit{EU} for manipulating the French people and damaging France's economy and sovereignty.

The second influential message group in the same convo represents another view of the influencers where they advocate for another candidate in the election: \textit{Florian Philippot} and support the \textit{LesPatriotes} party he leads. They also criticize president \textit{Macron} and call for a campaign against his policy on \textit{PassVaccinal}.

\paragraph{\#covid\_19 convo (Fig. \ref{fig:covid_clusters}):}
In both the influential message clusters of this convo, the influencers feel concerned about \textit{COVID\-19} and criticize \textit{Emmanuel Macron} for using the crisis for electoral gain and the improper policies. In the first message cluster, \textit{Vaccination passport} is supported and thought to be able to protect public health, while the \textit{Government} is accused for easing the \textit{Health constraints }for political purposes. Whereas in the second message cluster, people show desire for \textit{nightclubs}, advocate for \textit{Ivermectin} as a possible treatment for Covid-19, and think \textit{SudRadio} is a trustful resource of information and opinion.

\subsection{Analysis and Discussion}
To validate our hypotheses, we manually analyze the generated convo visualizations and influencer networks. In addition, we verify the generated agendas by comparing them with news articles and online forums. By this process we ascertain that they are correct representations of the broader corpus of data.

The numbers in Table \ref{tab:convo_stats} show that in both the convos, the top 10 influencers generate a very small fraction (5\% and 1\%) of the tweets in the convo, but receive a large proportion of retweets (44\% and 28\%). Therefore, we can reasonably extract the tweets by the top influencers to represent the influential or popular messages in the convo. A noticeable difference between the two convos is that, the number of the influential tweets in the \#Frexit convo is more than twice of the number in the \#covid\_19 convo. This signifies that the influencers in the \#Frexit convo have been much more active in generating content to drive their agendas.

We closely examine the network of influencers to analyze their retweet behavior. The\#frexit influencer network (Fig. \ref{fig:influencer_frexit}) is strongly connected by a large number of retweets. Bi-directional retweeting is also found between 4 pairs of authors. On the contrary, in the \#covid\_19 influencer network, the connections are notably weaker. All the connections are one-directional and majority of them originate from one author. We hypothesize that a combination of the influencers' efforts in generating the tweets and the density of their retweet network may serve as an early sign of a potential influence operation within the convo. Consequently, though the influencers' receive a substantial number of retweets in both convos, it is plausible that an influence operation may be more prominent in the \#Frexit convo compared to the \#covid\_19 convo.


Next, we analyze the agendas and convos snapshots generated by our model. In the \#Frexit convo, both the message clusters in Fig. \ref{fig:influencer_frexit} criticize president Macron but they advocate for different candidates \footnote{\url{https://www.thenationalnews.com/world/europe/2022/04/13/marine-le-pen-denies-frexit-agenda-as-presidential-rivals-clash-over-europe/}}. The first cluster focuses more on France's withdrawal from the EU, and support François Asselineau who is a lead character in the \#Frexit movement \footnote{\url{https://international.la-croix.com/news/politics/frances-frexit-presidential-candidate/4821}}. However, the second cluster advocates for Florian Philippot and his party, and shows strong opposition against the vaccine pass \footnote{\url{https://www.thelocal.fr/20180219/les-patriotes-florian-philippot-what-you-need-to-know-france-newest-far-right-party}}. Influencer \textit{I4} and \textit{I1} are the top contributors in the first cluster, while more than 90\% of the tweets in the second cluster come from \textit{I6}. Fig \ref{fig:influencer_frexit} also depicts that \textit{I4} and \textit{I1} retweet each other, and \textit{I6} is not directly connected to either of them. Hence, we can deduce that even within the same convo, distinct sub-groups of influencers and their respective retweeting users tend to agree on specific topics while disagreeing on others. Similarly, the agendas in the \#covid\_19 convo discuss several topics around the vaccine pass in France and criticize Macron for his handling of the pandemic \footnote{\url{https://www.france24.com/en/europe/20211014-macron-s-covid-health-pass-a-success-in-overcoming-france-s-vaccine-scepticism}}.

Therefore, these visualizations of the focused message groups in the convos provide a concise representation of the ongoing discussions among popular authors around that topic. Extracting such detailed analysis would require significant manual effort and time from experienced analysts. Together with the influencer networks, they provide valuable insights about targeted operations in a corpus.

\section{Related Work}

\subsection{Agenda and Emotion Detection}
In previous studies, agenda and emotion detection are usually presented as individual tasks.
Most of the work on agenda theory focus on the agenda setting process between the media and the audience, between different types of media, or between different groups of people \cite{ceron2016first, guo2019media, mccombs1997candidate, vargo2014network}. In \cite{mccombs1997candidate}, the authors investigate the images of the election candidates presented by mass media and shaped among the voters. Document level content analysis is performed on TV news and newspaper stories, and the voters' opinions towards the candidates are collected by surveys. 

The agendas in any political campaign are usually candidate names, their personalities, ideologies, and qualifications, and whether their images are built to be positive, neutral, or negative. The authors compare the proportion of each agenda between the media and voters to understand the effects of media on the general population.
With the emergence of social media, similar research has been performed to explore the agenda-setting dynamics between mass media and social media, to identify the roles of the agenda setter between them \cite{su2019agenda, conway2015agenda}.

The use of emotional language has also been an important indicator during any political campaign \citep{gruning2022emotional}.
With the availability of deep learning models and annotated datasets, fine-tuning models for emotion detection in particular domains have become a popular approach for emotion detecteion \citep{cai2018multi, huang2019emotionx, chiorrini2021emotion}. Prompting techniques using pre-trained large larguage models have also been deployed to perform zero-shot emotion detection in unknown domains \citep{plaza-del-arco-etal-2022-natural}. 
Although there has been significant research on emotion detection over individual messages on social media, our work is dedicated to detecting emotions expressed in groups of messages towards specific entities or topics.

All the works mentioned above aim to clarify the phenomena of information flow from one source to another, and ignore the possible influence operations behind the information flow. In contrast, our approach addresses this missing piece by identifying the influencers in the dynamic process and analyzing their behaviors and their inter-connections. Additionally, our break-down and summary method works efficiently as it requires lesser human annotation, which is a necessary step in most of the previous works \cite{guo2019media, su2022networked, subbiah2023agenda}.

\section{Conclusion}

In this work we introduce a novel systematic approach to identify influential message subsets from social media datasets. We build upon the idea of a \textit{convo} to extract concise snapshots of these online discussions around particular events or topics. We characterize these \textit{convos} using agendas and emotions to enable us to detect targeted influence operations by a group of influencers. We also create a network of these influencers to analyze the retweet behavior and group dynamics among them. A case study around prominent campaigns during the 2022 French Elections helps us to evaluate our proposed approach.
Future work would include application of our approach on other types of social media and definition of concrete metrics to measure degrees of influence and precisely identify influence operations.

\section{Acknowledgements}
This paper is based upon work supported by the Defense Advanced Research Projects Agency (DARPA) under Contract No. HR001121C0186. Any opinions, findings and conclusions or recommendations expressed in this material are those of the authors and do not necessarily reflect the views of DARPA or the U.S. Government.

\section{Limitations and Ethical Considerations}

Given the nature of the proposed task and the role of LLM in our approach, this work does have several limitations. Firstly, we analyze our approach on a publicly available Twitter dataset.
The dataset contains retweet information but does not differentiate between tweets, quotes, and replies. 
In future, we plan to extend to other platforms and other datasets with more diverse user interactions in order to better understand the group dynamics of online communities. 
Secondly, it is well known that the current large language models have a wide range of biases. Consequently, it is possible that our findings reflect the biased opinions nested in the LLM rather than from the dataset \citep{rozado2023political}.

We do not use or analyze any personally identifiable data in our experiments. The outputs generated are solely based on summaries generated by filtering popular messages on Twitter and the use of popular LLMs. They may reflect various political agendas or opinions of particular candidates. They must be verified by expert analysts before use in any downstream application.

\nocite{*}
\section{Bibliographical References}\label{sec:reference}

\bibliographystyle{lrec-coling2024-natbib}
\bibliography{custom}

\appendix
\section{Appendices}
\subsection{Appendix: Agenda and Emotion Detection Prompts}
\label{sec:appendix_a}

\begin{table*}[!t]
\begin{tabular}{p{0.95\textwidth}}
\hline
\small
\begin{verbatim}
[INST] <<SYS>>
You are a helpful, respectful and honest agenda detection assistant.
Read the list of messages given and understand the hidden agendas behind them.

Please do not share false information. If you are unable to understand the agenda,
simply say "No agenda".
<</SYS>>

What is the overall agenda behind this set of messages? Give me a short summary.
Messages: {input_text}
[/INST]
The agenda behind this set of tweets is
\end{verbatim}\\
\hline
\end{tabular}

\caption{System Prompt use to characterize convos using Llama-2-13b-chat}
\label{tab:prompt}
\vspace{1cm}

\begin{tabular}{p{0.95\textwidth}}
\hline
\small
\begin{verbatim}
What are the top distinct entities (maximum 5) mentioned in several messages?
What are the authors promoting about each entity? Give me 1 phrase for each.
Give the emotions expressed towards each entity.
Combine the entities, promoted actions and emotions in the following format:
output = [
{
"entity": {entity},
"promoted_actions": {action},
"emotions": {emotion}
},
...
]
\end{verbatim} \\
\hline
\end{tabular}
\caption{Prompts and output template to extract entities, promoted actions and emotions}
\label{tab:prompt}
\end{table*}

\end{document}